\documentclass[twoside,11pt]{article}

\usepackage{listings}
\usepackage{color}
\usepackage{graphicx}
\usepackage{enumerate}
\usepackage{enumitem}
\usepackage{caption}
\usepackage{jmlr2e}
\usepackage{subcaption}
\usepackage{hyperref}
\usepackage{amssymb}
\usepackage{amsmath}
\usepackage{listings}

\definecolor{purple}{rgb}{0.375,0,1}

\jmlrheading{18}{2017}{1-5}{9/15; Revised 10/16}{2/17}{15-492}{David Hallac, Christopher Wong, Steven Diamond, Abhijit Sharang, Rok Sosi\v{c}, Stephen Boyd, and Jure Leskovec}

\ShortHeadings{SnapVX: A Network-Based Convex Optimization Solver}{Hallac, Wong, Diamond, Sharang, Sosi\v{c}, Boyd, and Leskovec}
\firstpageno{1}

\begin{document}

\title{SnapVX: A Network-Based Convex Optimization Solver}

\author{\name David Hallac$^{*}$ \email hallac@stanford.edu \\
       \name Christopher Wong$^\dag$ \email crwong@cs.stanford.edu \\
       \name Steven Diamond$^\dag$ \email diamond@cs.stanford.edu \\
       \name Abhijit Sharang$^\dag$ \email abhisg@cs.stanford.edu \\
       \name Rok Sosi\v{c}$^\dag$ \email rok@cs.stanford.edu \\
       \name Stephen Boyd$^{*}$ \email boyd@stanford.edu \\
       \name Jure Leskovec$^\dag$ \email jure@cs.stanford.edu \\
      \addr Department of Electrical Engineering$^{*}$ \\
       Department of Computer Science$^\dag$ \\
       Stanford University, Stanford, CA, 94305}

\editor{Cheng Soon Ong}

\maketitle




\vspace{-2mm}
\begin{abstract}

SnapVX is a high-performance solver for convex optimization problems defined on networks. For problems of this form, SnapVX provides a fast and scalable solution with guaranteed global convergence.
It combines the capabilities of two open source software packages: Snap.py and CVXPY. Snap.py is a large scale graph processing library, and CVXPY provides a general modeling framework for small-scale subproblems.
SnapVX offers a customizable yet easy-to-use Python interface with ``out-of-the-box'' functionality. Based on the Alternating Direction Method of Multipliers (ADMM), it is able to efficiently store, analyze, parallelize, and solve large optimization problems from a variety of different applications. Documentation, examples, and more can be found on the SnapVX website at \url{http://snap.stanford.edu/snapvx}.

\end{abstract}


\section{Introduction}

Convex optimization is a widely used approach of modeling and solving problems in many different fields, as it offers well-established methods for finding globally optimal solutions. Numerous general-purpose optimization software packages exist \citep{sedumi, knitro, mosek, cvxpy}, but they typically rely on algorithms that are difficult to scale, so many modern machine learning problems cannot be solved by these common, yet general, approaches. Instead, solving large scale examples often requires developing problem-specific solution methods, which can be very fast but require significant optimization expertise to build. Furthermore, they are limited in scope, as they must be fine-tuned to only one particular type of problem and thus are hard to generalize.

In this paper, we build on the observation that many large convex optimization examples follow a common form in that they can often be split up into a series of subproblems using a network (graph) structure. Nodes are subproblems, representing anything from timestamps in a time-series data set to users in a social network. The edges then define the coupling, or relationships between the different nodes, and the combination of nodes and edges yields the original convex optimization problem
. This representation can refer to problems defined on actual networks, such as social or transportation systems, or questions classically modeled in other ways, such as control theory or time-series analysis.

Here, we present SnapVX, a solver that is both scalable and general on optimization problems defined over networks. It combines the graph capabilities of Snap.py \citep{snappy} with the general modeling framework from CVXPY \citep{cvxpy}. We show how SnapVX works, present syntax and supported features, scale it to large problems, and describe how it can be used to solve convex optimization problems from a variety of different fields. The full version of our solver, which is released under the BSD Open-Source License, can be found at the project website, \url{http://snap.stanford.edu/snapvx}. In addition to the source code, the download contains installation instructions, unit tests, documentation, and several examples to help users get started.

\section{SnapVX General Form}

Consider an optimization problem on an undirected graph $\mathcal{G} = (\mathcal{V}, \mathcal{E})$, with vertex set $\mathcal{V}$ and edge set $\mathcal{E}$, of the form:
\begin{equation}\label{original}
\begin{array}{ll} \underset{x}{\mathrm{minimize}} & \sum\limits_{i \in \mathcal V} f_i(x_i)
+ \sum\limits_{(j,k)\in \mathcal E} g_{jk}(x_j,x_k).
\end{array}
\end{equation}
Variable $x_i$ is associated with node $i$, for $i = 1, \ldots, |\mathcal{V}|$ (the size of $x_i$ can vary at each node). 
In problem \eqref{original}, $f_i
$
is the cost function at node $i$
, and $g_{jk}
$
is the cost associated with edge $(j,k)$. Constraints are represented by extended (infinite) values of $f_i$ and $g_{jk}$. Note that nodes and edges can also have local, private optimization variables (which can also vary in size). These remain 
confined to a single node or edge, though,
whereas the $x_i$'s are shared between different parts of the network. We consider only convex objectives and constraints for $f_i$ and $g_{jk}$, so the entire problem is a convex optimization problem.

To solve problem \eqref{original}, SnapVX uses a splitting algorithm based on the Alternating Direction Method of Multipliers (ADMM) \citep{BPCPE:11}. With ADMM, each individual component of the graph solves its own subproblem, iteratively passing small messages over the network and eventually converging to an optimal solution. See the SnapVX developer documentation for more details on the underlying derivation. By splitting up the problem into a series of subproblems, SnapVX is able to solve much larger convex optimization examples than standard solvers, which solve the whole problem at once. ADMM also keeps the optimality benefits that general solvers enjoy: not only are we guaranteed to obtain an optimal solution, but we also are given a certificate of optimality in the form of primal and dual residuals; see \citep{PB:14}.

SnapVX stores the network using a Snap.py graph structure. This allows fast traversal and easy manipulation, as well as efficient storage. On top of the standard graph, each node/edge is given convex objectives and constraints using CVXPY syntax. To find a global solution, SnapVX automatically splits up the problem and solves each subproblem using CVXPY, iteratively handling the ADMM message passing behind the scenes. Though wrapped in a Python layer, CVXPY uses ECOS and CVXOPT (two high-performance numerical optimization packages) as its primary underlying solvers \citep{cvxopt,ecos}, so the Python overhead is not significant and allows for easier interpretability and improved user interface. To parallelize the solver, a worker pool coordinates updates for the separate subproblems using Python's multiprocessing library. SnapVX is built to run on a single machine, parallelizing across multiple cores and allowing ``out-of-the-box'' functionality on machines ranging from standard laptops to large-memory servers.

\section{Syntax and Supported Features}
\label{sec:syntax}

We now present usage of SnapVX on a simple example. Complete documentation and more examples are available on the SnapVX website. Consider two nodes with an edge between them. We solve for a problem where each node has an unknown variable $x_i \in \mathbb{R}^1$. The first node's objective is to minimize $x_1^2 $ subject to $x_1 \leq 0$, the second's is to minimize $|x_2 + 3|$, and the edge objective penalizes the square norm difference between the two variables, $\|x_1 - x_2\|^2_2$. The following code specifies the optimization problem and solves it:

\lstset{language=Python}
\begin{lstlisting}[basicstyle=\small]
from snapvx import *
gvx = TGraphVX() #Create a new graph
x1 = Variable(1, name='x1') #Create a variable for node 1
gvx.AddNode(1, Objective=square(x1), Constraints=[x1 <= 0]) #Add new node
x2 = Variable(1, name='x2') #Repeat for node 2
gvx.AddNode(2, abs(x2 + 3), [])
gvx.AddEdge(1, 2, Objective=square(norm(x1 - x2)), Constraints=[]) #Add edge between nodes
gvx.Solve() #Solve the problem
gvx.PrintSolution() #Print the solution on a node-by-node basis
\end{lstlisting}

As SnapVX is meant to be a general-purpose solver, it has many customizable options to help easily and efficiently solve a wide range of convex optimization problems. These include ADMM iteration limits, verbose mode (to list intermediate steps at each iteration), and defining customized convergence thresholds. Two key features are highlighted below:

\begin{itemize}[nolistsep,leftmargin=.15in]
\item \textbf{Bulk Loading -} Often, the objectives and constraints at each node or edge will share a common form. For example, all the nodes could be trying to minimize $\|x - a_i\|_2$ for different values of $a_i$. 
Rather than requiring users to manually input each of these values, SnapVX allows for ``bulk loading'' of data. 
The functions \textit{AddNodeObjectives} and \textit{AddEdgeObjectives} allow the user to specify the general form of the objectives and an external file with separate data, and SnapVX fills in the details.
This functionality makes practical and significantly speeds up the loading of very large data sets.
\item \textbf{ADMM $\rho$-update -} The convergence time of ADMM depends on the value of the penalty parameter $\rho$ \citep{NLRPJ:15}, as it affects the tradeoff between primal and dual convergence, both of which need to be obtained for the overall problem to be solved. 
SnapVX users are not only able to select the value of $\rho$ (it defaults to $\rho = 1$), but can also define a function to update $\rho$ after each iteration based on the primal and dual residual values \citep{HYW:00, FB:15}.
\end{itemize}

\section{Scalability}

\begin{figure*}
\centering
\begin{minipage}{.5\textwidth}
  \centering
  \includegraphics[width=0.9\linewidth]{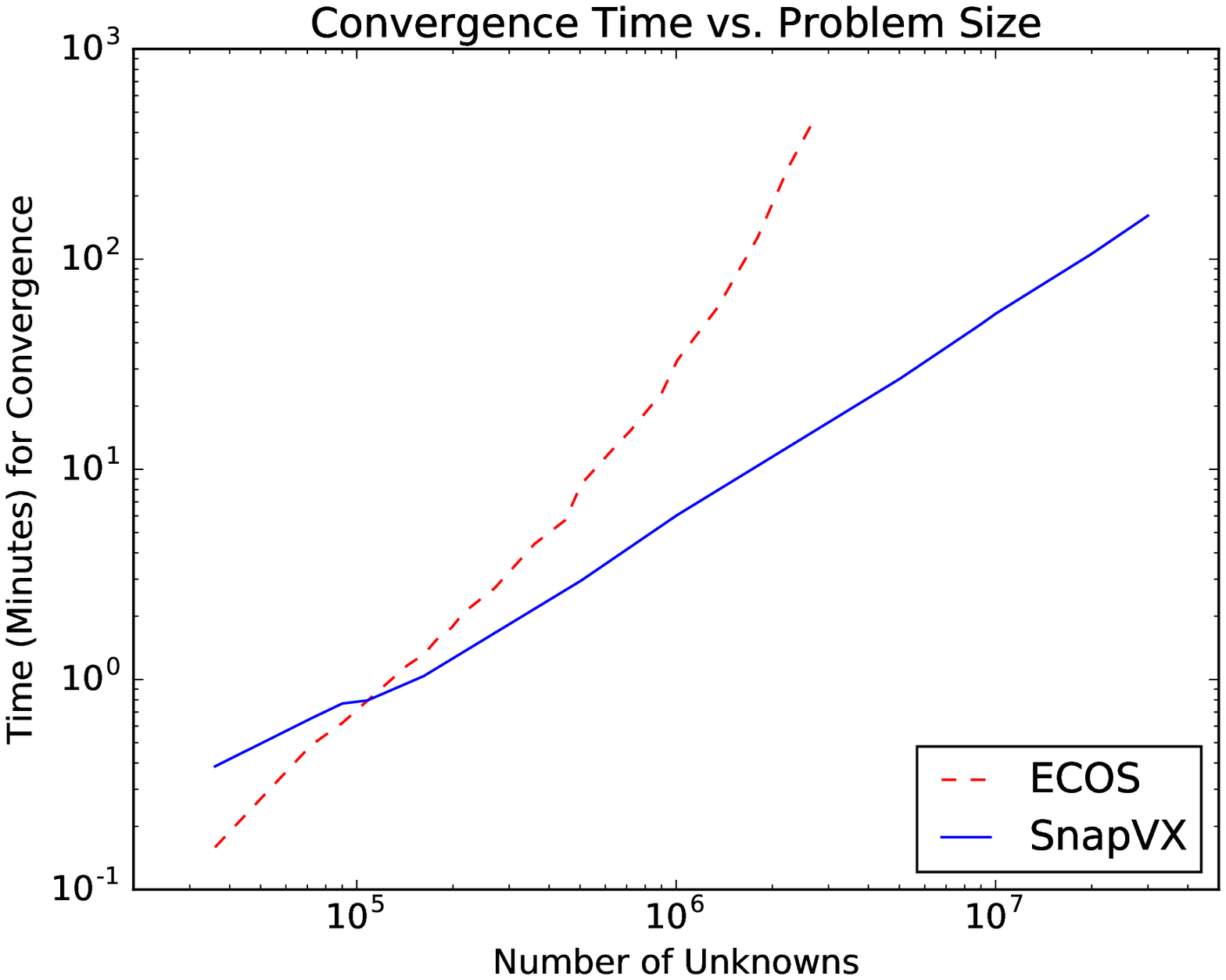}
\end{minipage}%
\begin{minipage}[t]{.51\textwidth}
  \centering
\scalebox{0.8}{
  \begin{tabular}{l r}
    \hline
    \# of Unknowns & SnapVX Solution Time (Minutes) \\ \hline 
    100,000 & 0.79\\ 
    500,000 & 2.95\\ 
    1 million & 6.06\\ 
    5 million & 27.30\\ 
    10 million & 55.42\\ 
    20 million & 106.33\\ 
    30 million & 161.95\\ \hline
    \end{tabular}
    }
\end{minipage}
\begin{minipage}{.45\textwidth}
  \centering
  \vspace{-5mm}
  \caption{Timing comparison of SnapVX and a general-purpose solver.}
  \label{graph1}
\end{minipage}%
\hskip 15pt%
\begin{minipage}{.45\textwidth}
  \centering
  \vspace{-5mm}
  \captionof{table}{SnapVX convergence time for several different problem sizes.}
  \label{table1}
\end{minipage}
\vspace{-7mm}
\end{figure*}

One of the biggest benefits of SnapVX is that it allows us to solve large problems very efficiently. 
It does so by automatically parallelizing the ADMM updates across multiple cores of a single machine using Python's multiprocessing class. Note that SnapVX is meant to be run on a single machine, rather than in a distributed computing environment.
Convergence time depends on the problem complexity, but we empirically observe that it scales approximately linearly with problem size. We compare SnapVX and a general solver for a problem on a 3-regular graph. Each node solves for an unknown variable in $\mathbb{R}^{9000}$, where the node objectives are sum-of-Huber penalties \citep{H:64} and edges have network lasso penalties \citep{HLB:15}.
By varying the number of nodes, we can span a wide range of problem sizes. The time required for SnapVX to converge, on a 40-core CPU where the entire problem can fit into memory, is shown in Table \ref{table1}. Figure \ref{graph1} displays a comparison to a general-purpose solver (ECOS, via CVXPY) in log-log scale. 

\vspace{-1mm}
\section{Applications}
SnapVX was created to allow users without significant optimization expertise to solve large network-based convex optimization problems by taking advantage of the powerful and scalable ADMM algorithm. 
Common examples from many different fields can be formulated in a SnapVX-friendly manner. Our software has been used to solve 
machine learning problems ranging from housing price prediction \citep{HLB:15} to ride-sharing analysis \citep{ghosh2016application} to high-dimensional regression \citep{yamada2016sparse}, often significantly outperforming other commonly used approaches. Examples in the SnapVX download package and on the project website --- including financial modeling, network inference, PageRank, and time-series analysis --- guide users towards applying the software to a variety of applications.
Both user and developer documentation help new users get started and, if they are interested, contribute to the code base. 
A robust set of unit tests ensures that the software has been properly downloaded and that the code is functioning as expected.
Overall, our open-source software provides an easy-to-use ADMM solver that can scale to large problems and apply to a wide variety of examples. With an active user base and rising interest from a range of scientific and engineering fields, we hope that SnapVX can become a useful tool for convex optimization problems, both in research and industry.

\vspace{-2mm}
\acks{This work was supported by NSF IIS-1149837, NIH BD2K, DARPA XDATA, DARPA SIMPLEX, Stanford Data Science Initiative, Boeing, Bosch, Lightspeed, SAP, and Volkswagen.}

\vskip 0.2in
\bibliography{refs}

\end{document}